\documentclass[12pt,a4paper]{article}
\usepackage{amssymb}
\usepackage{amsfonts}
\usepackage[onehalfspacing]{setspace}
\usepackage{geometry}

\newtheorem{theorem}{Theorem}

\newtheorem{corollary}[theorem]{Corollary}

\newtheorem{definition}{Definition}

\newtheorem{lemma}{Lemma}

\newtheorem{proposition}{Proposition}
\newtheorem{remark}[theorem]{Remark}

\input{tcilatex}
\begin{document}

\title{Machine-Learning to Trust\thanks{%
Financial support from UKRI Frontier Research Grant no. EP/Y033361/1 is
gratefully acknowledged. I thank Alex Clyde, Tuval Danenberg, Philippe
Jehiel, Nir Rosenfeld, Ariel Rubinstein, Yair Weiss, and referees of a
previously rejected version, for useful conversations and comments. Finally,
I acknowledge error-finding assistance from Refine.ink.}}
\author{Ran Spiegler\thanks{%
Tel Aviv University and University College London}}
\maketitle

\begin{abstract}
Can players sustain long-run trust when their equilibrium beliefs are shaped
by machine-learning methods that penalize complexity? I study a game in
which an infinite sequence of agents with one-period recall decide whether
to place trust in their immediate successor. The cost of trusting is
state-dependent. Each player's best response is based on a belief about
others' behavior, which is a coarse fit of the true population strategy with
respect to a partition of relevant contingencies. In equilibrium, this
partition minimizes the sum of the mean squared prediction error and a
complexity penalty proportional to its size. Relative to symmetric
mixed-strategy Nash equilibrium, this solution concept significantly narrows
the scope for trust.\bigskip \bigskip \pagebreak
\end{abstract}

\section{Introduction}

With the growing ubiquity of machine-learning (ML) algorithms, economists
are becoming increasingly interested in how ML\ affects economic
interactions. What makes this combination theoretically interesting is that
in contrast to traditional domains of ML, the training data that is likely
to serve ML in economic situations is often \textit{endogenous} --- i.e., it
is generated by the behavior of other agents, who may themselves be a
product of ML algorithms. Of particular interest are dynamic strategic
interactions, where players may use ML to learn how to respond to opponents'
history-dependent behavior --- which itself may be the product of
algorithmic learning. For example, as oligopolists adopt ML algorithms for
pricing decisions, a natural question of economic importance is how this
tendency impacts the scope for collusive pricing.

ML methods can be crudely classified into two paradigms, based on whether
the method involves constructing a predictive model of the environment
(Sutton and Barto (1998)). Model-free methods such as reinforcement learning
develop a direct association between actions and payoffs, without building a
predictive model of how actions map into payoff-relevant variables. In
contrast, model-based methods learn a predictive model of the environment's
data-generating process. The space of models that the algorithm explores may
be structured and interpretable (e.g., penalized linear regression or
probabilistic graphical models) or loose and difficult to interpret (e.g.,
random forests).

So far, the economic literature on dynamic strategic interactions between ML
algorithms has focused on the model-free variety, specifically on
reinforcement learning (see the literature review below). There have been no
attempts to construct models of long-run cooperation between human players
who act on beliefs shaped by model-based ML. And yet, given that predictive
AI is integrated with a growing variety of daily activities, it makes sense
to speculate how model-based ML might affect the scope for long-run
cooperation in dynamic interactions. In many settings, ML users will not be
satisfied with blunt action recommendations, because they will have an
internal or external need to be able to justify their behavior as a response
to a belief derived from a trusted source (which a model-based ML algorithm
may well be). This view is consistent with the observation in Agrawal et al.
(2002) that advances in AI \textquotedblleft decouple prediction and
judgment --- two primary ingredients to decision-making.\textquotedblright

Another reason that dynamic games between humans who use predictive,
model-based ML are interesting is that they raise a basic question about how
predictive ML models perform when their training data is endogenous. A
common feature of model-based ML is the penalty on complex predictive
models. The usual rationale behind it is trying to avoid over-fitting the
training data, which means lowering the variance of the algorithm's
out-of-sample predictions. In the context of dynamic strategic interactions,
this penalty on complexity may take the form of pooling different
contingencies (exogenous states of Nature, extensive-game histories) as if
they are equivalent.

In dynamic games, this force creates two sources of tension between what the
model-based-ML and individual-incentive perspectives regard as important
distinctions about opponents' play. First, for model-based ML, a rare event
is typically treated as unimportant, and the algorithm will tend to pool it
with other events to reduce model complexity. By comparison, from the
individual-incentive perspective, an event is important if it affects the
player's best-replying action, regardless of its frequency. For instance, in
the repeated Prisoner's Dilemma, when the threat of defection successfully
sustains cooperation, realization of this threat will be a rare event. Thus,
the two perspectives differ in how they link an event's importance to its
frequency.

Second, when opponents' behavior is similar in two contingencies,
model-based ML will tend to treat the two contingencies as equivalent for
predicting the opponent's future behavior. However, from the point of view
of individual incentives, two contingencies should be classified as
equivalent if they induce the same best-replying actions. Two beliefs can be
close for prediction purposes yet radically different in terms of the
best-reply they induce.

It follows that the model-based-ML and individual-incentive perspectives
have different notions of what makes two contingencies worth pooling. How
does this misalignment affect the prospect of long-run cooperation in
dynamic strategic interactions? To my knowledge, this is the first paper to
explore this question. Therefore, my approach here is to strive for maximum
simplicity: I devise the simplest possible game in which the above-described
tension can play out, and I use a stylized, caricature-like model of
model-based predictive ML.\bigskip

\noindent \textit{Preview of model and results}

\noindent I study the following discrete-time, infinite-horizon game. At
every period $t$, a distinct player chooses whether to put trust in his
immediate successor, player $t+1$. The cost of doing so is a random,
commonly observed state $\theta \in (0,1)$. Player $t$ derives a gross
benefit of $1$ if player $t+1$ chooses to put trust in his own successor,
player $t+2$. Players can condition their actions on the state and the most
recent action. This game is essentially an overlapping-generations \textit{%
trust game}, in which players' task is to predict how their immediate
successor's rate of cooperation will depend on their own action and the
payoff state. Symmetric mixed-strategy Nash equilibrium can sustain
arbitrarily high cooperation rates using a probabilistic Tit-for-Tat
strategy, such that at every payoff state, players are always indifferent
between cooperating and defecting.

I use this trust game as a \textit{template} for analyzing the broad
question of how ML-generated beliefs affect long-run cooperation. A useful
feature of this game is that the combination of random payoffs and reliance
on mixed strategies to sustain cooperation ensures that the statistical
tit-for-tat patterns that players' predictive ML models need to explain are
irreducibly rich. Thus, the trust game involves a repeated-game-like tension
between long-run mutual cooperation and short-run selfish motives, while at
the same time greatly simplifying the ML-based belief formation problem
without trivializing it.\footnote{%
Danenberg and Spiegler (2024) studied the OLG trust game to illustrate a
solution concept according to which players form beliefs by extrapolating
naively from representative finite samples drawn from the equilibrium
distribution.}

I adhere to the equilibrium modeling approach in analyzing the trust game,
while modifying the consistency criterion of equilibrium beliefs. The
modification captures in stylized form a common element of ML: explicitly
trading off a model's quality of empirical fit against its complexity. This
trade-off appears in various methods of supervised learning, e.g.
regularized regression (see Hastie et al. (2009)). Specifically, I assume
that players' belief is formed according to a partition of all relevant
contingencies (i.e., combinations of a payoff state and the recalled
history). The belief associated with a partition cell is the average
behavior in that cell, as in Jehiel (2005). The average is taken w.r.t the 
\textit{ergodic} \textit{distribution} over contingencies that is induced by
the population-level mixed strategy.

An \textit{ML-optimal} partition is required to \textit{minimize} the sum of
two terms: $(1)$ the Mean Squared Prediction Error (MSPE) of the
partition-induced belief, calculated according to the ergodic distribution;
and $(2)$ the partition's complexity, which is proportional to its size via
a constant $c>0$. A partition is strongly ML-optimal when it also satisfies
a refinement that essentially imposes continuity on the classification of
zero-probability contingencies. When the population-level mixed strategy
always assigns best-replies to beliefs induced by (strongly) ML-optimal
partitions, we have a \textit{(strong) ML equilibrium}.

The paper's main message is that the equilibrium criterion that penalizes
complex beliefs can drastically limit the scope for cooperation in the trust
game, for the two reasons mentioned above. Cooperative behavior relies on
the threat of a lower cooperation rate in response to defection. When the
cooperation rate is high, this threat is rarely realized. A
complexity-averse belief formation procedure may pool such a rare event with
more frequent ones, thus destroying the incentive to cooperate. The same
preference can also lead players to assign the same belief to different
contingencies at which the actual strategy prescribes similar behavior. Yet,
distinguishing between these contingencies may be crucial for maintaining
equilibrium incentives.

These forces generate several insights. First, perfect cooperation cannot be
sustained in equilibrium. In particular, if some states display full
cooperation, then the cooperation rate is bounded away from one in an equal
number of other states, regardless of how small $c$ is. Second, a positive
cooperation rate is impossible to sustain in equilibrium when the cost of
cooperation is $low$. Third, if $2cm^{3}>1$, it is impossible to sustain
cooperation in $m$ payoff states (or more). Thus, as the number of payoff
states grows, the cooperation rate that can be sustained in equilibrium
converges to zero. In this sense, trust between ML algorithms is harder to
achieve in complex environments. Fourth, if some payoff states exhibit no
cooperation at all, this may be lower the cooperation rate in other states.

Finally, I explore two variations on the model. First, I modify ML
optimality by requiring beliefs to be \textit{monotone} w.r.t underlying
payoffs (such that players always believe that the cooperation rate
decreases in the cost of cooperation). This acts as an additional impediment
to sustaining cooperation. Specifically, if the maximal cost of cooperation
across payoff states is $high$, then sustaining cooperation in all states is
impossible. In this sense, equilibrium trust among ML algorithms requires a
counter-intuitive belief in the relation between the rate of cooperation and
its cost. Second, I demonstrate via an example that duplicating actions is
not neutral in ML equilibrium, because players' history-dependent action
distributions have more room for spatial differentiation, thus curbing the
ML-optimality force that tends to bundle them in the same partition
cell.\bigskip

\noindent \textit{Related literature}

\noindent Eliaz and Spiegler (2019,2022) initiated the study of strategic
interactions with agents who form beliefs according to an ML-like procedure
that involves a fit-complexity trade-off. In these papers, a conventionally
rational agent reports his type to an automated agent who makes choices on
the agent's behalf, and uses penalized regression to form beliefs about the
appropriate action as a function of the agent's type.

This paper is also related to two strands in the behavioral game theory
literature. In one strand, Jehiel and Mohlin (2024) and Jehiel and Weber
(2024) take Jehiel's (2005) notion of Analogy-Based Expectations
Equilibrium, which captures strategic behavior under coarse beliefs, as a
starting point. They then apply basic ideas from the ML literature on
clustering to endogenize the analogy partitions that underlie players'
coarse equilibrium beliefs.

In Jehiel and Mohlin (2024), partition cells are shaped by an exogenous
notion of similarity between contingencies. However, they also respond to
the equilibrium frequency of contingencies --- in the spirit of the
bias-variance trade-off that is fundamental to the ML literature --- such
that infrequent contingencies are more likely to be grouped together.%
\footnote{%
Mohlin (2014) studies single-agent decision-making with endogenous formation
of coarse beliefs based on MSPE minimization.}

In Jehiel and Weber (2024), which is the closest precedent for MLEQ,
partition size is fixed at some $K$, and stability of partitions is
determined by (local or global) minimization of MSPE. The global version of
the solution concept in Jehiel and Weber (2024) can be viewed as a variant
of MLEQ, in which the cost of a partition is $0$ when its size is weakly
below $K$, and $\infty $ above it. The key departures in this paper from
Jehiel and Weber (2024) are the non-degenerate preference for simple
beliefs, and the dynamic context that endogenizes the distribution over
contingencies and gives rise to the above-described misalignment between
prediction and best-replying.

In the other strand, Spiegler (2002,2004,2005), Eliaz (2003), and Maenner
(2008) introduced the idea of complexity as an equilibrium belief-selection
criterion in dynamic games. They presented solution concepts in which an
Occam's-Razor-like principle rules out off-path threats as part of the
explanation of opponents' behavior. Unlike the present paper, these concepts
involved deterministic beliefs in strategies that take the form of finite
automata, in the tradition of Rubinstein (1986). As a result, the
ML-motivated trade-off between prediction error and complexity is moot in
these earlier works.

As mentioned earlier, the question of how the use of ML algorithms affects
collusive behavior in dynamic games has received recent attention. This
literature has focused on model-free, reinforcement-learning-based
algorithms; in this sense, it is orthogonal to the present paper. Therefore,
I make do with mentioning a handful of papers. Calvano et al. (2020) used
numerical experiments to demonstrate that a repeated oligopolistic pricing
game leads to collusive behavior by players who follow a
reinforcement-learning model known as Q-learning. Hansen et al. (2021),
Banchio and Mantegazza (2022) and Brown and Mackay (2023) extended the
numerical mode of analysis to other games, and also made progress in terms
of analytical characterizations. Waizmann (2024) studied an interaction
between a long-run player who obeys Q-learning and a sequence of rational
short-run players.\footnote{%
An older literature in evolutionary game theory examined repeated games when
players use reinforcement learning to adapt their actions over time ---
e.g., see Bendor et al. (2001).}

\section{A Model}

I begin by describing the trust game. Time is discrete and infinite: $%
t=0,1,2,3,...$. At each period $t$, a distinct player (also denoted $t$)
chooses an action $a_{t}\in \{0,1\}$. Player $0$ is a dummy whose action is
exogenously random. For every $t>1$, player $t$'s payoff only depends on his
action and the action taken by the subsequent player $t+1$. Specifically,
the payoff function is $u_{t}(a_{t},a_{t+1})=a_{t+1}-\theta a_{t}$, where $%
\theta $ is publicly observed and drawn uniformly from the set $\Theta
=\{\theta _{1},...,\theta _{n}\}\subset (0,1)$ at period $0$.

Each player $t>0$ observes $\theta $ and $a_{t-1}$ prior to taking his
action. I refer to $a_{t-1}$ as the \textit{observed history} at period $t$,
and use $h_{t}$ to denote it. Let $H=\{0,1\}$ denote the set of possible
observed histories. I refer to a pair $(\theta ,h)$ as a \textit{contingency}%
. Thus,\ every realization of $\theta $ induces an OLG game with complete
information and Prisoner's Dilemma payoffs, where players have one-period
recall.

I will be interested in symmetric strategy profiles in this game, where all
players $t>0$ obey the same mixed strategy $\sigma $, such that $\sigma
(a\mid \theta ,h_{t})$ is the (independent) probability that each player $%
t>0 $ plays $a$ in the contingency $(\theta ,h_{t})$. I adopt a population
interpretation of $\sigma $, and often use the \textit{shorthand notation} $%
\sigma (\theta ,h)$ for $\sigma (1\mid \theta ,h)$. Under the belief that
player $t+1$ follows a strategy $\hat{\sigma}$, player $t$'s expected payoff
from playing $a$ is $\hat{\sigma}(\theta ,a)-\theta a$. Therefore, $a=1$ ($0$%
) is a best-reply to $\hat{\sigma}$ if $\hat{\sigma}(\theta ,1)-\hat{\sigma}%
(\theta ,0)$ is weakly above (below) $\theta $.

For every payoff state $\theta $, a strategy $\sigma $ induces a \textit{%
two-state Markov process} over observed histories, where the probability of
transition from $h_{t}$ to $h_{t+1}$ is $\sigma (h_{t+1}\mid \theta ,h_{t})$%
. The long-run frequency of $a=1$ in the payoff state $\theta $ is the
invariant probability of $h=1$ induced by this Markov process. When $\sigma
(\theta ,0)>0$ or $\sigma (\theta ,1)<1$, this probability is well-defined
--- i.e., the Markov process has a well-defined invariant distribution $%
p_{\sigma }\in \Delta (\Theta \times H)$, given by%
\begin{equation}
p_{\sigma }(\theta ,1)=\frac{\sigma (\theta ,0)}{n[\sigma (\theta
,0)+1-\sigma (\theta ,1)]}  \label{ergodic}
\end{equation}%
and $p_{\sigma }(\theta ,0)=\frac{1}{n}-p_{\sigma }(\theta ,1)$. We say that 
$\sigma $ induces a \textit{positive cooperation rate} in $\theta $ if $%
p_{\sigma }(\theta ,1)>0$. When (\ref{ergodic}) is ill-defined, set $%
p_{\sigma }(\theta ,1)=1/2n$; this specification is only for completeness,
and has no effect on the results. The \textit{overall cooperation rate}
under $\sigma $ is $\sum_{\theta }p_{\sigma }(\theta ,1)$.\bigskip

\noindent \textit{Benchmark: Symmetric Nash equilibrium}

\noindent As usual, there is a Nash equilibrium in which players never
exhibit trust: $\sigma (\theta ,h)=0$ for every $(\theta ,h)$. Let us
explore other symmetric equilibria. Fix the strategy $\sigma $. If $\sigma
(\theta ,1)-\sigma (\theta ,0)>\theta $ ($<\theta $) for some $\theta $,
then any player's unique best-reply at $\theta $ is $a=1$ ($a=0$),
regardless of his observed history --- but this contradicts the optimality
of $\sigma (\theta ,0)<1$ ($\sigma (\theta ,1)>0$). If $\sigma (\theta
,1)-\sigma (\theta ,0)=\theta $, players are always indifferent between the
two actions, such that adhering to $\sigma $ is consistent with
best-replying.

It follows that any $\sigma $ that satisfies $\sigma (\theta ,1)-\sigma
(\theta ,0)=\theta $ for every $\theta $ is a symmetric Nash equilibrium
strategy. Note that since $\theta \in (0,1)$, $p_{\sigma }$ is well-defined.
In particular, we can set $\sigma (\theta ,1)=1$ and $\sigma (\theta
,0)=1-\theta $ for every $\theta $, such that the induced long-run
distribution $p_{\sigma }$ satisfies $p_{\sigma }(\theta ,1)=1/n$ for every $%
\theta $ --- i.e., players fully cooperate in equilibrium. $\blacksquare $%
\bigskip

Let us now introduce the novel belief-formation model, which draws
inspiration from the idea that beliefs are extrapolated from historical data
using ML methods. Fix the strategy $\sigma $ and its induced long-run
distribution $p_{\sigma }$. Let $\Pi $ be a partition of $\Theta \times H$.
Let $\pi (\theta ,h)$ denote the partition cell that includes $(\theta ,h)$.
Denote%
\[
p_{\sigma }(\pi )=\sum_{(\theta ,h)\in \pi }p_{\sigma }(\theta ,h) 
\]%
The \textit{representative strategy} of a cell $\pi \in \Pi $ with $%
p_{\sigma }(\pi )$ is%
\[
\hat{\sigma}(\pi )=\sum_{(\theta ,h)\in \pi }\frac{p_{\sigma }(\theta ,h)}{%
p_{\sigma }(\pi )}\sigma (\theta ,h) 
\]%
This is the expected strategy conditional on being in $\pi $, where the
expectation is taken w.r.t $p_{\sigma }$. When $\Pi $ is fixed and there is
no risk of confusion, I will use the notation $\hat{\sigma}(\theta ,h)$ as a
shorthand for $\hat{\sigma}(\pi (\theta ,h))$.

Define the function%
\[
V_{c,\sigma }(\Pi )=c\cdot \left\vert \Pi \right\vert +\sum_{(\theta
,h)}p_{\sigma }(\theta ,h)[\hat{\sigma}(\pi (\theta ,h))-\sigma (\theta
,h)]^{2} 
\]%
where $c>0$ is a constant capturing the cost of belief complexity. In the
spirit of machine classification algorithms, $V_{c,\sigma }$ trades off two
quantities: $(1)$ A classification's predictive accuracy, measured by the
representative strategy's \textit{mean squared prediction error} (\textbf{%
MSPE}); and $(2)$ the classification's complexity, measured by the partition
size.\bigskip

\begin{definition}[ML-optimality]
A partition $\Pi $ is \textit{ML-optimal} w.r.t $\sigma $ if it minimizes $%
V_{c,\sigma }(\Pi )$.\bigskip
\end{definition}

\begin{definition}[ML Equilibrium]
A strategy-partition pair $(\sigma ,\Pi )$ is an ML equilibrium if the
following two conditions hold:\smallskip \newline
$(i)$ $\Pi $ is ML-optimal w.r.t $\sigma $.\smallskip \newline
$(ii)$ if $\sigma (a\mid \theta ,h)>0$, then $a\in \arg \max_{a^{\prime }\in
\{0,1\}}[\hat{\sigma}(\pi (\theta ,a^{\prime }))-\theta a^{\prime }]$%
.\bigskip
\end{definition}

Thus, in ML Equilibrium (\textbf{MLEQ} in short), players' strategy
prescribes best-replies to a belief $\hat{\sigma}$, which in turn is induced
by an ML-optimal partition w.r.t players' strategy and the long-run
distribution over contingencies it induces.

If we assumed $c=0$, then the maximally fine partition would be ML-optimal,
such that we could have $\hat{\sigma}(\pi (\theta ,h))=\sigma (\theta ,h)$
for every $(\theta ,h)$ (including null ones), and MLEQ would collapse to
symmetric mixed-strategy Nash equilibrium.

The no-trust strategy $\sigma (\theta ,h)=0$ for every $(\theta ,h)$ is
consistent with MLEQ. To see why, note that under this $\sigma $, players
never vary their behavior with the contingency. As a result, a degenerate
partition of size $1$ induces zero MSPE, hence it is trivially ML-optimal.
This partition induces $\hat{\sigma}(\pi (\theta ,h))=0$ for every $(\theta
,h)$, such that players' unique best-reply is $a=0$, as postulated. Our task
will be to explore the possibility of SMLEQ with positive cooperation rates.

Note that if $\Pi $ is ML-optimal w.r.t $\sigma $, then $p_{\sigma }(\pi )>0$
for every $\pi \in \Pi $ --- otherwise, $\pi $ could be merged with some
other cell in $\Pi $, thus saving complexity costs without affecting
prediction errors.\bigskip 

\noindent \textit{A necessary condition for ML optimality}

\noindent The problem of finding ML-optimal partitions is computationally
complex (this is well-known in the clustering literature, which I discuss
below). Therefore, it is also unsurprising that exhaustive analytic
characterizations of ML-optimal partitions are hard to come by. Instead, I
will make repeated use of two necessary conditions for ML-optimal
partitions, which are themselves straightforward to check. The following
lemma states one of them.\bigskip

\begin{lemma}
\label{lemma merging}Suppose $\Pi $ is ML-optimal w.r.t $\sigma $. Then, the
following inequality holds for every two cells $\pi ,\pi ^{\prime }\in \Pi $:%
\begin{equation}
\frac{p_{\sigma }(\pi )p_{\sigma }(\pi ^{\prime })}{p_{\sigma }(\pi
)+p_{\sigma }(\pi ^{\prime })}\left( \hat{\sigma}(\pi )-\hat{\sigma}(\pi
^{\prime })\right) ^{2}\geq c  \label{ineq ML optimal}
\end{equation}%
\bigskip
\end{lemma}

The L.H.S of (\ref{ineq ML optimal}) represents the MSPE\ increase when we
deviate from $\Pi $ to a new partition that merges the cells $\pi $ and $\pi
^{\prime }$ into a single cell.\footnote{%
The formula is easy to derive using the variance decomposition formula. The
derivation appears in texts on clustering algorithms --- e.g., see Kaufman
and Rousseeuw (1990, pp. 230-231).} The R.H.S of (\ref{ineq ML optimal}) is
the complexity-cost reduction thanks to merging two cells. ML-optimality
requires the former to be weakly above the latter.\bigskip

\begin{corollary}
\label{corollary merging}Suppose $\Pi $ is ML-optimal w.r.t $\sigma $. Then, 
$\hat{\sigma}(\pi )\neq \hat{\sigma}(\pi ^{\prime })$ for every distinct $%
\pi ,\pi ^{\prime }\in \Pi $.\bigskip
\end{corollary}

This corollary immediately follows from Lemma \ref{lemma merging}. If $\hat{%
\sigma}(\pi )=\hat{\sigma}(\pi ^{\prime })$, we can merge the cells $\pi $
and $\pi ^{\prime }$ into one cell, thus lowering the partition's complexity
without changing its MSPE.\bigskip

\noindent \textit{Optimal assignment and Jehiel and Weber (2024)}

\noindent The definition of MLEQ is closely related to the notion of
Clustered Analogy-Based Expectations Equilibrium (Jehiel and Weber (2024)).
In both cases, players' equilibrium beliefs are extrapolated from the
objective distribution via an ML-inspired procedure. There are two
differences. First, Jehiel and Weber examine static games, where the
objective distribution over contingences is exogenous. In contrast, the
present game is dynamic, and the objective distribution over $\Theta \times
H $ is endogenously induced by the equilibrium strategy. Second, Jehiel and
Weber fix the partition size, whereas partition size in the present model is
variable and traded-off against prediction error.

Nevertheless, a simple observation by Jehiel and Weber (2024) provides the
second useful necessary condition for ML-optimal partitions: In an
ML-optimal partition, every contingency in the support of $p_{\sigma }$ is
assigned to a partition cell having the nearest representative
strategy.\bigskip

\begin{definition}[Optimal assignment]
A contingency $(\theta ,h)$ is optimally assigned w.r.t $(\sigma ,\Pi )$ if $%
\pi (\theta ,h)\in \arg \min_{\pi \in \Pi }\left\vert \hat{\sigma}(\pi
)-\sigma (\theta ,h)\right\vert $.\bigskip
\end{definition}

\begin{remark}
\label{lemma jehiel weber}Suppose that $\Pi $ is ML-optimal w.r.t $\sigma $.
If $p_{\sigma }(\theta ,h)>0$, then $(\theta ,h)$ is optimally assigned
w.r.t $(\sigma ,\Pi )$.\bigskip
\end{remark}

This is not a trivial observation, given that reassigning a contingency from
one cell to another may alter the cells' representative strategies. This
result is adapted from Lemma 1 in Jehiel and Weber (2024, Appendix B). Note
that while they can assume w.l.o.g that all contingencies have positive
probability, this is not guaranteed in the present context because $%
p_{\sigma }$ is endogenous. Motivated by this observation, I present the
following definition, which extends the optimal assignment property to
zero-probability contingencies.\bigskip

\begin{definition}[Strong ML-optimality]
A partition $\Pi $ is strongly ML-optimal w.r.t $\sigma $ if it is
ML-optimal w.r.t $\sigma $, and if every contingency is optimally assigned
w.r.t $(\Pi ,\sigma )$.\bigskip
\end{definition}

\begin{definition}[Strong MLEQ]
A MLEQ $(\sigma ,\Pi )$ is strong if $\Pi $ is strongly ML-optimal w.r.t $%
\sigma $.\bigskip
\end{definition}

\noindent In what follows, I use the abbreviation \textbf{SMLEQ} to describe
a strong MLEQ.\footnote{%
In Jehiel and Weber (2024), optimal assignment arises in a variant on their
solution concept, in which partitions need not minimize MSPE, but
contingencies are required to be optimally assigned. Such partitions can be
obtained via a simple iterative procedure due to Lloyd (1975).}

The optimal assignment property implies the following property of
SMLEQ.\bigskip

\begin{remark}
\label{remark tit for tat}Let $(\sigma ,\Pi )$ be a SMLEQ. Then, for every $%
\theta $, $\sigma (\theta ,1)\geq \sigma (\theta ,0)$ and $\hat{\sigma}%
(\theta ,1)\geq \hat{\sigma}(\theta ,0)$.\bigskip
\end{remark}

Thus, the intuitive (weak) tit-for-tat property that characterizes symmetric
Nash equilibrium carries over to SMLEQ: Cooperative behavior leads to weakly
higher cooperation rate, both objectively and according to players'
beliefs.\bigskip

\noindent \textit{Discussion}

\noindent Several features of the model merit discussion. First, mixed
strategies are indispensable for sustaining cooperation in symmetric Nash
equilibrium. Yet, once we admit them, we can sustain arbitrary cooperation
rates in each state. As we will see, the indifference conditions that
characterize equilibrium with mixed strategies will play a key role in our
analysis of MLEQ, without producing the \textquotedblleft anything
goes\textquotedblright\ aspect of symmetric Nash equilibrium.

Symmetric Nash equilibrium satisfies local stability w.r.t a population
dynamic process, in which at every period a small subset of the population
of players gets to adjust their behavior, and in addition there is an
invasion by small group of mutants who play the pure tit-for-tat strategy.
The reason is that the invasion slightly increases the population-level
difference $\sigma (\theta ,1)-\sigma (\theta ,0)$ above $\theta $. Since $%
a=1$ is the unique best-reply to this difference, regardless of the history,
the adjusting agents in the population will lower the population-level
difference $\sigma (\theta ,1)-\sigma (\theta ,0)$ back toward $\theta $.
This local-stability argument does not work for other types of mutations.

Second, the notion of ML-optimality imposes a penalty on complex beliefs,
even though the MSPE is calculated w.r.t the \textit{actual} long-run
distribution $p_{\sigma }$ over contingencies. In practice, the ML rationale
for penalizing complexity is that empirical fit is calculated against a
finite sample; penalizing complexity mitigates data over-fitting and thus
improves out-of-sample predictions. When the sample is infinite (which is
implicitly the case in our model), this rationale vanishes. Therefore, the
appropriate way to interpret simplicity seeking in MLEQ is that it is a
tractable \textit{reduced form} of a more elaborate model in which players
extrapolate beliefs from finite samples drawn from the endogenous
distribution.\footnote{%
Danenberg and Spiegler (2024) is an example of such a more elaborate model,
albeit without the crucial simplicity-seeking component.} I elaborate on
this argument in Appendix II.

MLEQ also has a \textquotedblleft behavioral\textquotedblright\
interpretation, according to which the complexity penalty captures an aspect
of \textit{human} players' departure from the standard Bayesian-rational
model. The idea is that human players are guided by an Occam's-razor-like
principle that favors simple explanations of their opponents' behavior (as
in Spiegler (2002,2004,2005) and Eliaz (2003)). This interpretation follows
the tradition of Simon (1956) and Rubinstein (1998), where elements of AI
serve as metaphors for procedural human rationality.

Third, ML-optimality pools together contingencies that involve different
payoff states, rather than handling each state in isolation --- even though
when players make decisions, they are informed of $\theta $. The
interpretation is that the ML algorithms which the notion of ML-optimality
approximates gather behavioral data without having an \textquotedblleft
understanding\textquotedblright\ of the deep incentive structure that
generates the data. The payoff state $\theta $ may itself be determined by
multiple observable variables; even if the payoff implications of these
variables are directly perceived by the player when he makes his choices,
they need not be transparent to the ML algorithm. Furthermore, recall that
the informal story underlying ML-optimality is that the algorithm operates
on a finite sample. Even if the algorithm \textquotedblleft
understands\textquotedblright\ the need to distinguish between payoff
states, it may lack sufficient data to do so without sacrificing predictive
success.

This \textquotedblleft lack of understanding\textquotedblright\ that
ML-optimality captures has two specific implications. First, ML-optimal
partitions need not preserve the product structure of the set of
contingencies. For example, $(\theta ,1)$ and $(\theta ^{\prime },1)$ can be
pooled into the same cell, while $(\theta ,0)$ and $(\theta ^{\prime },0)$
belong to different cells. Second, ML-optimal partitions may produce beliefs
that are not monotone in $\theta $, hence they may be hard to interpret in
light of the game's underlying incentives.

\section{Examples}

In this section I present two examples that illustrate MLEQ and demonstrate
how the penalty on complex beliefs, inherent in the notion of ML-optimality,
constrains the ability to sustain trust in MLEQ.

\subsection{Example I: $n=1$}

This is the simplest specification of the model. When there is only one cost
value $\theta \in (0,1)$, there are two possible contingencies, hence a
partition size of $2$ induces rational expectations, as the partition
isolates each contingency in a separate cell. In contrast, when the
partition size is $1$, $\hat{\sigma}(1)=\hat{\sigma}(0)$ (because $\theta $
is the only state, I do not include it as an explicit argument of $\sigma $, 
$\hat{\sigma}$ and $p_{\sigma }$). Since $\hat{\sigma}(1)-\hat{\sigma}%
(0)<\theta $, the only best-reply to $\hat{\sigma}$ is $a=0$. It follows
that the only way to sustain a positive cooperation rate in MLEQ is with the
fine, size-$2$ partition. In this case, $\hat{\sigma}(h)=\sigma (h)$ for
every $h$, such that in a trust-sustaining MLEQ, $\sigma (1)-\sigma
(0)=\theta $, as in Nash equilibrium. The only additional requirement is
that the fine partition is ML-optimal, which in this case is equivalent to
the inequality (\ref{ineq ML optimal}). Since $p_{\sigma }(1)+p_{\sigma
}(0)=1$, the inequality is reduced to%
\[
p_{\sigma }(1)p_{\sigma }(0)\cdot (\sigma (1)-\sigma (0))^{2}\geq c
\]%
Plugging the formula for $p_{\sigma }(h)$ and the equilibrium indifference
equation $\sigma (1)-\sigma (0)=\theta $, we obtain the following condition
for $\sigma $ to be consistent with equilibrium:%
\begin{equation}
\frac{(\sigma (1)-\theta )(1-\sigma (1))}{(1-\theta )^{2}}\theta ^{2}\geq c
\label{example n=1 condition}
\end{equation}

It can be easily checked that if $c>\theta ^{2}/4$, there exists no solution
to (\ref{example n=1 condition}). Thus, a necessary and sufficient condition
for the existence of trust-sustaining MLEQ is $c\leq \theta ^{2}/4$. Thus,
perhaps surprisingly, cooperation is harder to sustain when the cost of
cooperation $\theta $ is \textit{low}.

When $c\leq \theta ^{2}/4$, the maximally cooperative equilibrium strategy
is given by the value of $\sigma (1)\geq (1+\theta )/2$ that solves (\ref%
{example n=1 condition}) bindingly. Importantly, this value is bounded away
from $1$. In particular, when $c=\theta ^{2}/4$, $\sigma (1)=(1+\theta )/2$
such that $p_{\sigma }(1)=\frac{1}{2}$.

Thus, the penalty on complex beliefs implies a limit on the amount of trust
that can be sustained in MLEQ. The reason is that a high cooperation rate
implies that $p_{\sigma }(0)$ is low, such that the simplicity-seeking
motive pushes ML-optimal beliefs to pool $h=0$ together with the other
contingency $h=1$, thereby destroying the incentive to cooperate.

\subsection{Example II: $n=2$}

The following result characterizes the maximal amount of cooperation that
can be sustained in MLEQ when there are two payoff states, under certain
parametric restrictions. The proof of this result (like all subsequent
proofs) is relegated to Appendix I.\bigskip

\begin{proposition}
\label{prop n=2}Let $\theta _{2},\theta _{1}>\frac{1}{2}$ and $c\in (\frac{1%
}{8},\frac{1}{4})$. Any MLEQ that exhibits a positive cooperation rate in
some state $\theta $ must satisfy $\left\vert \Pi \right\vert =2$; $%
p_{\sigma }(\theta ,1)\leq \theta ^{2}/(1+\theta ^{2})$; and $p_{\sigma
}(\theta ^{\prime },1)=0$ in $\theta ^{\prime }\neq \theta $. If $\theta
^{2}<\sqrt{c}/(1-\sqrt{c})$, cooperation in state $\theta $ is
unsustainable. If $\theta ^{2}\geq \sqrt{c}/(1-\sqrt{c})$, the upper bound
on $p_{\sigma }(\theta ,1)$ is sustainable.\bigskip
\end{proposition}

The derivation does not take a stand on whether $\theta $ is $\theta _{1}$
or $\theta _{2}$. Since $\theta _{2}>\theta _{1}$, we can see that the
largest possible equilibrium probability of $h=1$ in any state is $\theta
_{2}^{2}/(1+\theta _{2}^{2})$. Thus, a lower cost of cooperation leads to a
lower upper bound on the long-run probability of cooperative behavior, and
also makes it harder to sustain cooperation at all.

The partition that sustains cooperation in state $\theta $ (under the
parametric restrictions in the statement of the result) isolates $(\theta
,1) $ in a separate cell, and pools the three other contingencies in another
cell. Since $(\theta ^{\prime },1)$ and $(\theta ^{\prime },0)$ are in the
same cell, there are no incentives to cooperate in state $\theta ^{\prime }$%
. As a result, $\sigma (\theta ^{\prime },1)=\sigma (\theta ^{\prime },0)=0$
in equilibrium. This lowers the average cooperation rate in the cell that
includes these contingencies, which is also $\hat{\sigma}(\theta ,0)$. The
indifference condition $\hat{\sigma}(\theta ,1)-\hat{\sigma}(\theta
,0)=\theta $ thus means that $\hat{\sigma}(\theta ,1)$ is also pulled down.
It follows that \textit{failure to cooperation in one state exerts a
negative equilibrium effect on the cooperation rate in the other state}.

In the $n=1$ example, we saw that the only possible deviation in the $\Pi $
dimension was merging the contingencies $h=1$ and $h=0$ into one cell, thus
destroying the incentive to cooperate. When $n=2$, we should also consider a
deviation that merges the contingencies $(\theta _{1},h)$ and $(\theta
_{2},h)$ into one cell, thus potentially destroying a distinction between
the two payoff states that is important for sustaining the incentive to
cooperate. Proposition \ref{prop n=2} reflects this additional force. As a
result, the scope for cooperation is lower than in the $n=1$ case, where it
is possible to sustain $p_{\sigma }(1)\geq \frac{1}{4}$ whenever $\theta
^{2}\geq 4c$. In the proof of Proposition \ref{prop n=2}, I present
inequality (\ref{nec condition n=2}), a necessary condition for sustaining
cooperation when $n=2$, which is more stringent than the inequality $\theta
^{2}\geq 4c$.

\section{General Results}

The first result establishes a necessary condition for SMLEQ to sustain
positive cooperation rates in multiple payoff states. For a fixed complexity
cost, there is an upper bound on the number of payoff states in which
cooperation can be sustained in SMLEQ. The bound is not tight, because it is
calculated without taking into account the endogeneity of $p_{\sigma }$%
.\bigskip

\begin{proposition}
\label{prop bound}Let $m$ be integer satisfying $2cm^{3}>1$. Then, for
generic $\Theta $, there exists no SMLEQ that induces positive cooperation
rates in $m$ payoff states (or more).\bigskip
\end{proposition}

The logic behind the result is as follows. First, in order to sustain
cooperation in $m$ payoff states, players' beliefs must be sufficiently
complex to create the distinctions that sustain cooperation incentives.
Specifically, I show that for generic $\Theta $, the equilibrium partition
must have at least $m+1$ cells. Second, as a partition gets larger, the
probability of individual cells goes down on average, and so does the
average distance between the cells' average cooperation rate. These two
quantities are inversely proportional to $m+1$ and $(m+1)^{2}$,
respectively. As $m$ grows larger, a deviation in belief space that merges
two cells of a putative equilibrium partition becomes more likely.\bigskip

\begin{corollary}
Fix $c$. As $n\rightarrow \infty $, the overall cooperation rate in SMLEQ
for generic $\Theta $\ converges to zero.\bigskip
\end{corollary}

This result immediately follow from Proposition \ref{prop bound}. Since the
maximal number of payoff states with positive cooperation rates is bounded
from above by $\sqrt[3]{1/2c}$, the fraction of these states becomes
negligible as $n$ grows larger.

The following result generalizes the observation in Section 3 that low costs
of cooperation are an obstacle to sustaining cooperation in SMLEQ.\bigskip

\begin{proposition}
\label{prop max theta}Let $m$ be an integer. For generic $\Theta $, if $%
m\cdot \max (\Theta )<\sqrt{2c}$, then there is no SMLEQ that induces
positive cooperation rates in $m$ payoff states (or more).\bigskip
\end{proposition}

Thus, when we fix $c$ and $m$ and lower the range of values of $\theta $, it
becomes harder to sustain cooperation in equilibrium. The bound on $\max
(\Theta )$ in this result is quite inefficient, but it enables a proof that
is a simple variant on the proof of Proposition \ref{prop bound}. As in the
example of Section 3.2, the intuition is that if cooperation costs are
small, the indifference conditions that characterize equilibrium behavior
imply that the $\hat{\sigma}$'s are relatively close to each other, which
makes inequality (\ref{ineq ML optimal}) harder to satisfy.

The following result exploits the endogeneity of $p_{\sigma }$ to derive a
different kind of limit on the ability to sustain cooperation in
MLEQ.\bigskip

\begin{proposition}
\label{prop 2m payoff states}Suppose that under a SMLEQ, $p_{\sigma }(\theta
,1)=\frac{1}{n}$ in $m$ payoff states. Then, for generic $\Theta $, there
are $m$ additional payoff states in which the cooperation rate is bounded
away from $1$ for any $c>0$.\bigskip
\end{proposition}

Thus, if we wish to sustain full cooperation in SMLEQ in $m$ payoff states,
then the cooperation rate is lower than one --- by an amount that is bounded
away from zero --- even if we let the complexity cost vanish. In particular,
Proposition \ref{prop 2m payoff states} implies that the fraction of payoff
states that exhibit full cooperation (i.e., $p_{\sigma }(\theta ,1)=\frac{1}{%
n}$) is at most $\frac{1}{2}$.

The logic behind this result is as follows. If $\sigma (\theta ,1)=1$ for
some $\theta $, then $p_{\sigma }(\theta ,0)=0$. SMLEQ then requires the
equilibrium partition to bundle $(\theta ,0)$ with other
non-zero-probability contingencies, which in turn means that there is some
other payoff state for which the cooperation rate is below $1$. An
additional argument establishes that this pairing must be different for
different payoff states, hence for every state that exhibits full
cooperation there is a distinct state in which the cooperation rate is
bounded away from one.

\section{Two Variations}

In this section I modify the model in two orthogonal directions. First, I
redefine MLEQ such that the domain of possible predictive models is
restricted in a way that is intuitive a priori. Second, I show that unlike
Nash equilibrium, duplicating actions is not neutral under MLEQ.

\subsection{Monotone Beliefs}

A key feature of ML-optimal partitions is that they reflect no preconception
about the structure of $\hat{\sigma}(\theta ,h)$. As we saw, equilibrium
conditions imply that $\hat{\sigma}(\theta ,1)\geq \hat{\sigma}(\theta ,0)$
for every $\theta $, which is monotone in an intuitive, tit-for-tat
direction. However, equilibrium conditions do not imply any systematic, let
alone monotone relation between $\hat{\sigma}$ and $\theta $. In this
section I examine a variant on the model, in which $\hat{\sigma}$ is
required to be weakly decreasing in $\theta $, reflecting a preconception
that a higher cost of cooperation should lead to a lower propensity to
cooperate.\bigskip

\begin{definition}[Monotone ML optimality]
A partition $\Pi $ is monotone w.r.t $\sigma $ if $\hat{\sigma}(\theta
,h)\geq \hat{\sigma}(\theta ^{\prime },h)$ for every $h$ and $\theta
^{\prime }>\theta $. A partition is monotone ML-optimal w.r.t $\sigma $ if
it minimizes $V_{c,\sigma }(\Pi )$ among all partitions that are monotone
w.r.t $\sigma $.\bigskip
\end{definition}

\begin{definition}[Monotone ML equilibrium]
A pair $(\sigma ,\Pi )$ is a monotone ML equilibrium if: $(i)$ $\Pi $ is
monotone ML-optimal w.r.t $\sigma $; and $(ii)$ if $\sigma (a\mid \theta
,h)>0$, then $a$ is a best-reply to $\hat{\sigma}(\pi (\theta ,h))$.\bigskip
\end{definition}

Monotonicity is $not$ an impediment for cooperation in symmetric Nash
equilibrium: The strategy $\sigma $ defined by $\sigma (\theta ,1)=1$ and $%
\sigma (\theta ,0)=1-\theta $ for every $\theta $ is weakly decreasing in $%
\theta $, and it constitutes a symmetric Nash equilibrium that sustains full
cooperation. In contrast, the following result shows a limitation on
sustaining cooperation in monotone MLEQ.\bigskip

\begin{proposition}
\label{prop monotone MLEQ}For every $c>0$, there exists $\delta <1$ such
that if $\max (\Theta )>\delta $, there is no monotone MLEQ that sustains
positive cooperation rates in every payoff state.\bigskip 
\end{proposition}

Thus, in order to sustain positive cooperation rates in all payoff states
(including ones with a high cost of cooperation), ML-optimal beliefs must be
counter-intuitive, in the sense that expected cooperation rates cannot be
monotonically decreasing in the cost of cooperation. In this sense, the
\textquotedblleft lack of understanding\textquotedblright\ that the basic
form of MLEQ exhibits is necessary for it to function well in the trust game.

Proposition \ref{prop monotone MLEQ} crucially relies on the monotonicity
requirement. Without it, even when $\max (\Theta )$ is arbitrarily close to
one, it may be possible to construct a (non-monotone) SMLEQ partition that
separates contingencies of the form $(\theta ,1)$ while grouping together
contingencies of the form $(\theta ,0)$. As a result, even if $\sigma
(\theta ,1)$ is high such that the long-run probability of $(\theta ,0)$ is
low, the grouping together of low-frequency contingencies of the latter form
can ensure that no further complexity-reducing merging of partition cells
will take place.

Note that monotone MLEQ imposes monotone ML optimality not just on the
equilibrium partition, but also on candidate deviation partitions. This
restricts the set of deviations relative to a version that only impose
monotonicity on the equilibrium partition. In this sense, that alternative
version would make it \textit{even harder} to sustain cooperation relative
to our notion of monotone ML equilibrium.

\subsection{Duplicating Actions}

In the trust game, players' prediction problem involves a scalar variable:
the probability that the successor will play $a=1$. In a more general game
with an arbitrary number of actions, the object of players' prediction
problem is a probability vector. Would this higher dimensionality have
fundamental effects on MLEQ? Thorough exploration of this question is beyond
the scope of the present paper. However, I will use a simple example to
argue that the answer is affirmative: The mere increase in the
dimensionality of predicted objects can give history-contingent beliefs
\textquotedblleft more room\textquotedblright\ to keep away from each other,
thus weakening the complexity-reducing force that underlies ML\ optimality.

Let us modify Example I from Section 3.1, by assuming that players' action
space is $A=\{0,0^{\ast },1,1^{\ast }\}$, where $0^{\ast }$ and $1^{\ast }$
are payoff-equivalent to $0$ and $1$, respectively. Such action duplication
is of course irrelevant for Nash equilibrium analysis. However, we will see
it is not MLEQ-neutral. A strategy $\sigma $ assigns an action mixture $%
\sigma (\cdot \mid h)\in \Delta (A)$ to every observed history $h\in A$.

Because I am about to compare MLEQ across action spaces of different
cardinality, I need to redefine the prediction-error criterion.
Specifically, I use the squared Hellinger distance (see Basu et al. (2011))
as a measure of the distance between two action mixtures $\sigma $ and $%
\sigma ^{\prime }$:%
\[
H^{2}(\sigma ^{\prime },\sigma )=\frac{1}{2}\dsum\limits_{a\in A}\left( 
\sqrt{\sigma ^{\prime }(a)}-\sqrt{\sigma (a)}\right) ^{2} 
\]%
This measure has the advantage that for given $\sigma $ and $\sigma ^{\prime
}$, if we duplicate actions and split $\sigma $ and $\sigma ^{\prime }$
uniformly into the duplicates, $H^{2}(\hat{\sigma},\sigma )$ remains
unchanged. Thus, the action-duplication exercise will not generate effects
that mechanically rely on an inappropriate measure of prediction error.

Accordingly, I assume that given a partition cell $\pi $, the representative
strategy is%
\[
\hat{\sigma}(a\mid \pi )=\left( \sum_{(\theta ,h)\in \pi }\frac{p_{\sigma
}(\theta ,h)}{p_{\sigma }(\pi )}\sqrt{\sigma (a\mid \theta ,h)}\right) ^{2} 
\]%
for every $a\in A$. A partition $\Pi $ is evaluated by%
\[
V_{c,\sigma }(\Pi )=c\cdot \left\vert \Pi \right\vert +\frac{1}{2}%
\sum_{(\theta ,h)}p_{\sigma }(\theta ,h)\dsum\limits_{a\in A}\left( \sqrt{%
\hat{\sigma}(a\mid \pi (\theta ,h))}-\sqrt{\sigma (a\mid (\theta ,h))}%
\right) ^{2} 
\]

Under these modified definitions, the inequality that sustains cooperation
in MLEQ in the original one-state, two-action example (i.e., when $A=\{0,1\}$%
) is as follows (I omit any reference to $\theta $ from $\sigma $ and $p$,
because there is only one state):%
\begin{equation}
p_{\sigma }(1)p_{\sigma }(0)[1-\sqrt{\sigma (1)\sigma (0)}-\sqrt{(1-\sigma
(1))(1-\sigma (0))}]\geq c  \label{ineq duplicate section}
\end{equation}

Now consider the duplicated-actions case, $A=\{0,0^{\ast },1,1^{\ast }\}$.
Construct the following strategy $\sigma $:%
\begin{eqnarray*}
\sigma (a &=&1\mid h=1)=\sigma (a=1\mid h=1^{\ast })=\alpha \\
\sigma (a &=&0\mid h=1)=\sigma (a=0\mid h=1^{\ast })=1-\alpha \\
\sigma (a &=&1^{\ast }\mid h=0)=\sigma (a=1^{\ast }\mid h=0^{\ast })=\beta \\
\sigma (a &=&0^{\ast }\mid h=0)=\sigma (a=0^{\ast }\mid h=0^{\ast })=1-\beta
\end{eqnarray*}%
Under $\sigma $, players mix between the original actions $1$ and $0$ after $%
h\in \{1,1^{\ast }\}$, and between the duplicate actions $1^{\ast }$ and $%
0^{\ast }$ after $h\in \{0,0^{\ast }\}$. By definition, ML optimality
requires $\pi (1)=\pi (1^{\ast })=\pi $ and $\pi (0)=\pi (0^{\ast })=\pi
^{\prime }$, such that $\hat{\sigma}(\pi )=(1-\alpha ,0,\alpha ,0)$ and $%
\hat{\sigma}(\pi ^{\prime })=(0,1-\beta ,0,\beta )$. The indifference
condition requires $\alpha -\beta =\theta $. Moreover, the invariant
probabilities of the two cells, $p_{\sigma }(\pi )$ and $p_{\sigma }(\pi
^{\prime })$, are equal to their original two-action analogues, $p_{\sigma
}(1)$ and $p_{\sigma }(0)$, when $\sigma (1)=\alpha $ and $\sigma (0)=\beta $%
. The only difference is in the term that the prediction error contributes
to the inequality:%
\[
\frac{1}{2}[(\sqrt{1-\alpha })^{2}+(\sqrt{1-\beta })^{2}+(\sqrt{\alpha }%
)^{2}+(\sqrt{\beta })^{2}]=1 
\]%
which is greater than the corresponding term $1-\sqrt{\sigma (1)\sigma (0)}-%
\sqrt{(1-\sigma (1))(1-\sigma (0))}$ in (\ref{ineq duplicate section}).

Thus, what is essentially the same strategy as in the original one-state,
two-action example becomes easier to sustain in MLEQ under action
duplication. The reason is that now, the distance between $\hat{\sigma}(\pi
) $ and $\hat{\sigma}(\pi ^{\prime })$ is larger than in the two-action
analogue. Although the Hellinger-distance criterion rules out mechanical
changes in distance due to the higher dimensionality of $A$, the
construction of $\sigma $ enables history-dependent beliefs to be spaced
more widely apart: The action mixtures in response to cooperation and
defection live in different dimensions. This mitigates the ML-optimality
force that tends to merge cells. It follows that the extension to multiple
actions can create more opportunities for sustaining long-run cooperation in
MLEQ.

\section{Conclusion}

The ML dilemma between what I described here as model-based vs. model-free
methods has occupied AI specialists (e.g., see Sutton and Barto (1998) and
Levine et al. (2020)), and involves technical considerations that lie very
far outside my expertise. Nevertheless, to the extent that ML methods are
applied to dynamic strategic decision-making, I hope that this paper has
made a valuable contribution to the discussion.

Model-based ML involves reducing the dimensionality of the environment's
representation. In game-theoretic contexts, this may take the form of
collapsing distinct contingencies into the same equivalence class. The basic
qualitative insight of this paper is that when this classification is guided
by weighing predictive success against model complexity, it may fail to make
the equilibrium distinctions that sustain individual incentives to act
cooperatively in a long-run interaction. Successful long-run cooperation is
sustained by counterfactual (or at least rare) threats, yet their very
rarity causes ML methods that involve a trade-off between empirical fit and
simplicity to group them together with other contingencies. This same
trade-off causes similar beliefs to be grouped together, even when they
imply radically different best-replying actions.

This paper used the OLG trust game as a convenient template for a systematic
exploration of these themes, and demonstrated that they can drastically
narrow the scope for equilibrium long-run cooperation. More specifically, we
saw that cooperation becomes harder to sustain when the cost of cooperation
is low, when there are many payoff states, and when the ML method is
constrained to induce intuitively monotone beliefs. We also saw that
cooperation patterns in some payoff states constrain the ability to sustain
cooperation in others. These lessons are easily extendible to arbitrary
two-action OLG games: Symmetric Nash equilibria that rely on off-path
threats must involve mixing, which generates essentially the same
considerations as in the trust game, and would therefore give rise to
similar results.

\noindent {\LARGE Appendix I: Proofs}\bigskip

\noindent {\large Remark \ref{remark tit for tat}}

\noindent Let $(\sigma ,\Pi )$ be a SMLEQ. Consider a payoff state $\theta $%
. Suppose $\hat{\sigma}(\theta ,1)-\hat{\sigma}(\theta ,0)\neq \theta $.
Then, players have the same, unique best-replying action at $(\theta ,1)$
and $(\theta ,0)$, such that $\sigma (\theta ,1)=\sigma (\theta ,0)$. The
optimal assignment property implies that $\pi (\theta ,1)=\pi (\theta ,0)$,
hence $\sigma (\theta ,1)=\sigma (\theta ,0)=0$. Now suppose $\hat{\sigma}%
(\theta ,1)-\hat{\sigma}(\theta ,0)=\theta $. Then, $\hat{\sigma}(\theta ,1)>%
\hat{\sigma}(\theta ,0)$. If $\sigma (\theta ,1)<\sigma (\theta ,0)$, this
violates the optimal assignment property. $\blacksquare $\bigskip

\noindent {\large Proposition \ref{prop n=2}}

\noindent As before, a SMLEQ $(\sigma ,\Pi )$ can sustain trust with
positive probability only if $\left\vert \Pi \right\vert >1$. Let us
distinguish between two cases.\bigskip

\noindent \textit{Case 1: }$\left\vert \Pi \right\vert \geq 3$.

\noindent\ W.l.o.g, assume $\hat{\sigma}(\theta _{1},1)-\hat{\sigma}(\theta
_{1},0)=\theta _{1}$. Since $\theta _{1},\theta _{2}>\frac{1}{2}$, $\hat{%
\sigma}(\theta _{1},0)<\frac{1}{2}$. If players exhibit trust with positive
probability in $\theta _{2}$, then $\hat{\sigma}(\theta _{2},0)<\frac{1}{2}$
as well; otherwise, $\hat{\sigma}(\theta _{2},0)=0$. In either case, $%
\left\vert \hat{\sigma}(\theta _{1},0)-\hat{\sigma}(\theta
_{2},0)\right\vert <\frac{1}{2}$.

Let us consider two sub-cases. First, suppose $\hat{\sigma}(\theta
_{1},0)\neq \hat{\sigma}(\theta _{2},0)$, such that $(\theta _{1},0)$ and $%
(\theta _{2},0)$ belong to different partition cells. Note that $\alpha
\beta /(\alpha +\beta )<\frac{1}{2}$ for every $\alpha ,\beta \in \lbrack
0,1]$. It follows that%
\[
\frac{p_{\sigma }(\pi (\theta _{1},0))p_{\sigma }(\pi (\theta _{2},0))}{%
p_{\sigma }(\pi (\theta _{1},0))+p_{\sigma }(\pi (\theta _{2},0))}\left( 
\hat{\sigma}(\theta _{1},0)-\hat{\sigma}(\theta _{2},0)\right) ^{2}<\frac{1}{%
8}<c 
\]%
in violation of (\ref{ineq ML optimal}). Second, suppose $\hat{\sigma}%
(\theta _{1},0)=\hat{\sigma}(\theta _{2},0)$. Then,%
\begin{eqnarray*}
\left\vert \hat{\sigma}(\theta _{2},1)-\hat{\sigma}(\theta
_{1},1)\right\vert &=&\left\vert \hat{\sigma}(\theta _{2},0)+\theta _{2}-%
\hat{\sigma}(\theta _{1},0)-\theta _{1}\right\vert \\
&=&\left\vert \theta _{2}-\theta _{1}\right\vert \\
&<&\frac{1}{2}
\end{eqnarray*}%
such that%
\[
\frac{p_{\sigma }(\pi (\theta _{1},1))p_{\sigma }(\pi (\theta _{2},1))}{%
p_{\sigma }(\pi (\theta _{1},1))+p_{\sigma }(\pi (\theta _{2},1))}\left( 
\hat{\sigma}(\theta _{1},1)-\hat{\sigma}(\theta _{2},1)\right) ^{2}<\frac{1}{%
8}<c 
\]%
in violation of (\ref{ineq ML optimal}). $\square $\bigskip

\noindent \textit{Case 2: }$\left\vert \Pi \right\vert =2$.

\noindent We will first rule out partitions that consist of two equally
sized cells:\bigskip

\noindent $(i)$ Suppose $\Pi $ consists of the cells $\{(\theta
_{1},0),(\theta _{2},0)\}$ and $\{(\theta _{1},1),(\theta _{2},1)\}$. Then, $%
\hat{\sigma}(\theta _{1},h)=\hat{\sigma}(\theta _{2},h)$ for both $h=0,1$.
However, this is inconsistent with equilibrium. To see why, note first that
if $\hat{\sigma}(\theta ,1)-\hat{\sigma}(\theta ,0)>\theta $ for some $%
\theta $, then we must have in equilibrium that $\sigma (\theta ,1)=\sigma
(\theta ,0)=1$. But then, by Remark \ref{lemma jehiel weber}, both $(\theta
,1)$ and $(\theta ,0)$ should be assigned to the cell with the highest $\hat{%
\sigma}$, a contradiction. In the same manner, we can rule out the
possibility that $\hat{\sigma}(\theta ,1)-\hat{\sigma}(\theta ,0)<\theta $
for some $\theta $. It follows that $\hat{\sigma}(\theta ,1)-\hat{\sigma}%
(\theta ,0)=\theta $ for both $\theta $, a contradiction. Thus, we can rule
out this particular partition.\medskip

\noindent $(ii)$ Now suppose $\Pi $ consists of the cells $\{(\theta
_{1},1),(\theta _{1},0)\}$ and $\{(\theta _{2},1),(\theta _{2},0)\}$. Then, $%
\hat{\sigma}(\theta ,1)=\hat{\sigma}(\theta ,0)$ for both $\theta $. The
unique best-reply against $\hat{\sigma}$ is to play $a=0$, contradicting the
assumption that the equilibrium sustains trust.\medskip

\noindent $(iii)$ Finally, the partition that consists of $\{(\theta
_{1},1),(\theta _{2},0)\}$ and $\{(\theta _{2},1),(\theta _{1},0)\}$ implies 
$\hat{\sigma}(\theta ,0)>\hat{\sigma}(\theta ,1)$ for some $\theta $, thus
making $a=0$ the unique best-reply to $\hat{\sigma}$ at $\theta $. But then $%
\sigma (\theta ,1)=\sigma (\theta ,0)=0$, contradicting the assignment of $%
(\theta ,1)$ and $(\theta ,0)$ to different cells.\bigskip

It follows that if $\Pi $ is consistent with MLEQ, one of its two cells must
be a singleton $\{(\theta ,h)\}$. This means that for $\theta ^{\prime }\neq
\theta $, $(\theta ^{\prime },1)$ and $(\theta ^{\prime },0)$ are assigned
to the same cell, which (as we have seen) means that in equilibrium, $\sigma
(\theta ^{\prime },1)=\sigma (\theta ^{\prime },0)=0$. Suppose that this
cell also contains $(\theta ,1)$, such that the other, singleton cell is $%
\{(\theta ,0)\}$. It follows that $\hat{\sigma}(\theta ,0)=\sigma (\theta
,0) $. Equilibrium requires that $\hat{\sigma}(\theta ,1)-\hat{\sigma}%
(\theta ,0)=\theta $, hence $\hat{\sigma}(\theta ,1)>\hat{\sigma}(\theta ,0)$%
. By Remark \ref{lemma jehiel weber}, ML-optimality requires that since $%
\sigma (\theta ^{\prime },1)=\sigma (\theta ^{\prime },0)=0<\hat{\sigma}%
(\theta ,0)$, $(\theta ^{\prime },0)$ and $(\theta ^{\prime },1)$ should be
bundled together with $(\theta ,0)$ rather than $(\theta ,1)$, a
contradiction.

The only remaining option is that $\Pi $ consists of two cells that take the
form $\{(\theta ^{\prime },1),(\theta ^{\prime },0),(\theta ,0)\}$ and $%
\{(\theta ,1)\}$, such that 
\begin{eqnarray*}
\hat{\sigma}(\theta ,1) &=&\sigma (\theta ,1) \\
\hat{\sigma}(\theta ,0) &=&\frac{p_{\sigma }(\theta ,0)}{\frac{1}{2}%
+p_{\sigma }(\theta ,0)}\sigma (\theta ,0)
\end{eqnarray*}%
Equilibrium requires $\sigma (\theta ,1)-\hat{\sigma}(\theta ,0)=\theta $.
ML-optimality requires that if we move $(\theta ,0)$ out of its cell and
into the same cell as $(\theta ,1)$, the mean squared error will not
decrease:%
\begin{equation}
\frac{\frac{1}{2}\cdot p_{\sigma }(\theta ,0)}{\frac{1}{2}+p_{\sigma
}(\theta ,0)}(\sigma (\theta ,0)-0)^{2}\leq \frac{p_{\sigma }(\theta
,0)\cdot p_{\sigma }(\theta ,1)}{p_{\sigma }(\theta ,0)+p_{\sigma }(\theta
,1)}(\sigma (\theta ,1)-\sigma (\theta ,0))^{2}  \label{ineq example k=2}
\end{equation}%
\bigskip Plugging the expressions for $p_{\sigma }(\theta ,h)$ and the
equilibrium requirement $\sigma (\theta ,1)-\hat{\sigma}(\theta ,0)=\theta $
into (\ref{ineq example k=2}), and with a bit of algebra, we can derive the
following tight upper bound:%
\[
p_{\sigma }(\theta ,1)\leq \frac{\theta ^{2}}{1+\theta ^{2}} 
\]%
This inequality is binding when (\ref{ineq example k=2}) is binding.

The only remaining constraint is that the putative equilibrium is robust to
deviating from the two-cell partition to the degenerate partition:%
\[
\frac{p_{\sigma }(\pi (\theta ,1))p_{\sigma }(\pi (\theta ,0))}{p_{\sigma
}(\pi (\theta ,1))+p_{\sigma }(\pi (\theta ,0))}(\hat{\sigma}(\theta ,1)-%
\hat{\sigma}(\theta ,0))^{2}=p_{\sigma }(\theta ,1)(1-p_{\sigma }(\theta
,1))\theta ^{2}\geq c 
\]%
Since $p_{\sigma }(\theta ,1)<\frac{1}{2}$, the L.H.S of this inequality
holds only if it holds under the upper bound on $p_{\sigma }(\theta ,1)$,
which happens if and only if%
\begin{equation}
c\leq \frac{\theta ^{4}}{(1+\theta ^{2})^{2}}  \label{nec condition n=2}
\end{equation}%
This inequality is equivalent to $\theta ^{2}\geq \sqrt{c}/(1-\sqrt{c})$. $%
\blacksquare $\bigskip

\noindent {\large Proposition \ref{prop bound}}

\noindent Consider an SMLEQ $(\sigma ,\Pi )$ in which $p_{\sigma }(\theta
,1)>0$ for $m$ payoff states $\theta $. The proof proceeds stepwise.\bigskip

\noindent \textbf{Step 1}: \textit{If }$p_{\sigma }(\theta ,1)>0$\textit{,
then }$\hat{\sigma}(\theta ,1)-\hat{\sigma}(\theta ,0)=\theta $\textit{%
.\smallskip }

\noindent \textbf{Proof}: Consider a payoff state $\theta $ for which $%
p_{\sigma }(\theta ,1)>0$. Suppose $\hat{\sigma}(\theta ,1)-\hat{\sigma}%
(\theta ,0)<\theta $. Then, the unique best-reply at $\theta $ is $a=0$.
Hence, $\sigma (\theta ,1)=\sigma (\theta ,0)=0$ in equilibrium,
contradicting the assumption that $p_{\sigma }(\theta ,1)>0$. Now suppose $%
\hat{\sigma}(\theta ,1)-\hat{\sigma}(\theta ,0)>\theta $. Then, the unique
best-reply at $\theta $ is $a=1$. Hence, $\sigma (\theta ,1)=\sigma (\theta
,0)=1$ in equilibrium. By the optimal assignment property of SMLEQ, $(\theta
,1)$ and $(\theta ,0)$ must both be assigned to $\arg \max_{\pi \in \Pi }%
\hat{\sigma}(\pi )$, such that $\hat{\sigma}(\theta ,1)=\hat{\sigma}(\theta
,0)$, a contradiction. $\square $\bigskip

\noindent \textbf{Step 2}: \textit{For generic }$\Theta $\textit{, if }$%
p_{\sigma }(\theta ,1)>0$\textit{\ for }$m$\textit{\ payoff states }$\theta $%
\textit{, then }$\left\vert \Pi \right\vert >m$\textit{.\smallskip }

\noindent \textbf{Proof}: Consider a payoff state $\theta $ for which $%
p_{\sigma }(\theta ,1)>0$. By Step 1, $\hat{\sigma}(\theta ,1)\neq \hat{%
\sigma}(\theta ,0)$, which by Corollary \ref{corollary merging} means that $%
\pi (\theta ,1)\neq \pi (\theta ,0)$. Construct a \textit{non-directed graph}
whose nodes correspond to the cells in $\Pi $, such that $\pi $ and $\pi
^{\prime }$ are linked if there is $\theta $ for which $p_{\sigma }(\theta
,1)>0$ such that $\pi (\theta ,h)=\pi $ and $\pi (\theta ,1-h)=\pi ^{\prime
} $ for some $h$. Note that by definition, the graph has $m$ edges.

Suppose $\pi (\theta ,1)$ and $\pi (\theta ,0)$ are linked. Then, by Step 1, 
$\hat{\sigma}(\theta ,1)-\hat{\sigma}(\theta ,0)=\theta $. Suppose the graph
contains an additional, \textit{indirect} path between $\pi (\theta ,1)$ and 
$\pi (\theta ,0)$. By Step 1, this means that there is a sequence of payoff
states $\theta ^{1},...,\theta ^{K}$, such that $\sum_{k=1}^{K}\theta
^{k}=\theta $. For generic $\Theta $, this requirement fails to hold. It
follows that if two graph nodes are linked, there is no additional indirect
path between them (in other words, the link is a bridge). It follows that
the graph must be a forest (i.e., every connected graph component is a
tree), hence it has at least $m+1$ nodes. It follows that $\left\vert \Pi
\right\vert >m$. Note also that each connected component in this forest
consists of at least two nodes. $\square $\bigskip

\noindent \textbf{Step 3}: \textit{Formulating an auxiliary max-min
problem\smallskip }

\noindent \textbf{Proof}: We have established that $\Pi $ consists of $K\geq
m+1$ cells, each with its own distinct $\hat{\sigma}$. Enumerate the
partition cells as $\pi _{1},...,\pi _{K}$. Denote $p_{i}=p_{\sigma }(\pi
_{i})$ and $\hat{\sigma}_{i}=\hat{\sigma}(\pi _{i})$. By (\ref{ineq ML
optimal}), $(\sigma ,\Pi )$ is an MLEQ only if%
\begin{equation}
c\leq \min_{i\neq j}\frac{p_{i}p_{j}}{p_{i}+p_{j}}\left( \hat{\sigma}_{i}-%
\hat{\sigma}_{j}\right) ^{2}  \label{necessary condition step 3}
\end{equation}%
for every distinct $i,j\in \{1,...,K\}$. By definition, the R.H.S. of this
inequality is bounded from above by%
\begin{equation}
\max_{p\in \Delta \{1,...,K\},\text{ }\hat{\sigma}\in \lbrack
0,1]^{K}}\min_{i\neq j}\frac{p_{i}p_{j}}{p_{i}+p_{j}}\left( \hat{\sigma}_{i}-%
\hat{\sigma}_{j}\right) ^{2}  \label{max min expression}
\end{equation}

Without loss of generality, let $0\leq \hat{\sigma}_{1}<\hat{\sigma}%
_{2}<\cdots <\hat{\sigma}_{K}\leq 1$. For every $k=1,...,K-1$, denote $q_{k}=%
\hat{\sigma}_{k+1}-\hat{\sigma}_{k}$. Denote $p=(p_{k})_{k=1,...,K}$ and $%
q=(q_{k})_{k=1,...,K-1}$. By definition, (\ref{max min expression}) is
weakly below%
\begin{equation}
\max_{p,q}\min_{k=1,...,K-1}\frac{p_{k}p_{k+1}}{p_{k}+p_{k+1}}q_{k}^{2}
\label{max min 2nd expression}
\end{equation}%
By definition, $p\in \Delta \{1,...,K\}$ is a probability $n$-vector,
whereas $q_{k}>0$ for every $k=1,...,K-1$ and $\sum_{k}q_{k}\leq 1$. Since (%
\ref{max min 2nd expression}) is increasing in $q$, we can take (as far as
the solution of this max-min problem is concerned) the latter constraint to
be binding, such that $q\in \Delta \{1,...,K-1\}$. $\square $\bigskip

\noindent \textbf{Step 4:} \textit{The value of (\ref{max min 2nd expression}%
) is strictly below }$1/2(K-1)^{3}$\textit{.\smallskip }

\noindent \textbf{Proof:} Let us break the max-min problem into two steps:%
\[
\max_{p}\left( \max_{q}\min_{k=1,...,K-1}\frac{p_{k}p_{k+1}}{p_{k}+p_{k+1}}%
q_{k}^{2}\right) 
\]%
As a first step, fix $p$. For every $k=1,...,K-1$, denote 
\[
A_{k}=\sqrt{\frac{p_{k}\,p_{k+1}}{p_{k}+p_{k+1}}} 
\]%
Since $k$ is selected to minimize $(A_{k}q_{k})^{2}$, it is clear that the
solution to the parenthetical max-min problem $\max_{q}%
\min_{k}(A_{k}q_{k})^{2}$ equalizes $A_{k}q_{k}$ across all $k$, such that 
\[
q_{k}=\frac{\frac{1}{A_{k}}}{\sum_{j=1}^{K-1}\frac{1}{A_{j}}} 
\]%
and the max-min value is%
\begin{equation}
\frac{1}{\left( \sum_{j=1}^{K-1}\frac{1}{A_{j}}\right) ^{2}}
\label{maxmin value}
\end{equation}

In the procedure's second step, choose $p$ to maximize this value. This is
equivalent to choosing $p$ to minimize%
\begin{equation}
\sum_{k=1}^{K-1}\sqrt{\frac{1}{p_{k}}+\frac{1}{p_{k+1}}}
\label{intermediate objective}
\end{equation}%
I will now derive a (non-tight) lower bound for the minimum of (\ref%
{intermediate objective}). Note that (\ref{intermediate objective}) is
strictly convex. It is also symmetric between $p_{1}$ and $p_{K}$, as well
as across all interior components $p_{2},...,p_{K-1}$. Therefore, the
expression's unique minimizer has full support and satisfies $p_{1}=p_{K}$. 

Construct an alternative probability vector $p^{\ast }=(p_{1},...,p_{K-1})$
defined as follows:%
\[
p_{k}^{\ast }=\frac{p_{k}}{\sum_{j=1}^{K-1}p_{j}}
\]%
Then, we can rewrite (\ref{intermediate objective}) as%
\[
\frac{1}{\sqrt{\sum_{j=1}^{K-1}p_{j}}}\left( \sum_{k=1}^{K-2}\sqrt{\frac{1}{%
p_{k}^{\ast }}+\frac{1}{p_{k+1}^{\ast }}}+\sqrt{\frac{1}{p_{K-1}^{\ast }}+%
\frac{1}{p_{1}^{\ast }}}\right) 
\]%
which is strictly greater than%
\[
\sum_{k=1}^{K-2}\sqrt{\frac{1}{p_{k}^{\ast }}+\frac{1}{p_{k+1}^{\ast }}}+%
\sqrt{\frac{1}{p_{K-1}^{\ast }}+\frac{1}{p_{1}^{\ast }}}
\]%
The latter expression is strictly convex and symmetric. Therefore, its
unique minimizer is the uniform distribution, yielding a minimal value of $%
(K-1)\sqrt{2(K-1)}$. It follows that (\ref{maxmin value}) is strictly below $%
1/2(K-1)^{3}$. $\square $\bigskip 

\noindent \textbf{Step 5:} \textit{Completing the proof\smallskip }

\noindent It follows from Steps 3 and 4 that (\ref{max min expression}), and
therefore also $c$, are below $1/2(K-1)^{3}$. Since $K\geq m+1$, this
contradicts the assumption that $2cm^{3}>1$. We can conclude that there is
no SMLEQ with positive cooperation rates in $m$ payoff states. $\blacksquare 
$\bigskip

\noindent {\large Proposition \ref{prop max theta}}

\noindent Steps 1 and 2 in the proof of Proposition \ref{prop bound} apply
here. Consider the necessary condition given by the inequality (\ref%
{necessary condition step 3}). Since $p_{i}p_{j}/(p_{i}+p_{j})\leq \frac{1}{2%
}$, the condition implies%
\[
c\leq \min_{i\neq j}\frac{1}{2}\left( \hat{\sigma}_{i}-\hat{\sigma}%
_{j}\right) ^{2} 
\]%
Recall that the auxiliary graph constructed in Step 2 of the proof of
Proposition \ref{prop bound} is a forest consisting of $K>m$ nodes, where
each component consists of at least two nodes. Consider an arbitrary
connected component of the forest, consisting of $L>1$ nodes. For every pair 
$i,j$ of linked nodes in this component, $\left\vert \hat{\sigma}_{i}-\hat{%
\sigma}_{j}\right\vert \in \Theta $. Therefore, for every $i,j$ in this
component, $\hat{\sigma}_{i}-\hat{\sigma}_{j}\leq (L-1)\cdot \max (\Theta )$%
. The necessary condition for equilibrium thus implies%
\[
c\leq \frac{1}{2}(L-1)\cdot \max (\Theta ))^{2}\leq \frac{1}{2}(m\cdot \max
(\Theta ))^{2} 
\]%
in violation of the assumption that $m\cdot \max (\Theta )<\sqrt{2c}$. $%
\blacksquare $\bigskip

\noindent {\large Proposition \ref{prop 2m payoff states}}

\noindent Denote $\Theta ^{\ast }=\{\theta \in \Theta \mid p_{\sigma
}(\theta ,1)=\frac{1}{n}\}$. Then, $\sigma (\theta ,1)=1$ for every $\theta
\in \Theta ^{\ast }$. This means that an ML-optimal partition must group all
contingencies $(\theta ,1)=1$ with $\theta \in \Theta ^{\ast }$ in the same
cell $\pi ^{\ast }$. Since $p_{\sigma }(\theta ,0)=0$ for every $\theta \in
\Theta ^{\ast }$, SMLEQ requires that every contingency $(\theta ,0)$ with $%
\theta \in \Theta ^{\ast }$ is grouped with some other contingency in a cell 
$\pi _{\theta }\neq \pi ^{\ast }$, such that $\hat{\sigma}(\pi ^{\ast })-%
\hat{\sigma}(\pi _{\theta })=\theta $. By definition, $\pi _{\theta }$ must
contain at least one no-null contingency $(\theta ^{\prime },h)\neq (\theta
,0)$, such that $\sigma (\theta ^{\prime },h)\leq \hat{\sigma}(\pi _{\theta
})$.

Let us consider two cases. First, suppose $h=1$. By the optimal assignment
property, $\sigma (\theta ^{\prime },1)$ is weakly closer to $\hat{\sigma}%
(\pi _{\theta })$ than to $\hat{\sigma}(\pi ^{\ast })$. Therefore, $\sigma
(\theta ^{\prime },1)\leq (1+1-\theta )/2$. By Remark \ref{remark tit for
tat}, $\sigma (\theta ^{\prime },0)\leq (\theta ^{\prime },1)$. Therefore, $%
p_{\sigma }(\theta ^{\prime },1)\leq 1-\theta /2$, and therefore bounded
away from $1$, where the gap is independent of $c$.

Second, suppose $h=0$. Therefore, if the cooperation rate in $\theta
^{\prime }$ is positive, then $\hat{\sigma}(\theta ^{\prime },1)=\hat{\sigma}%
(\pi _{\theta })+\theta ^{\prime }$. Moreover, $\pi (\theta ^{\prime
},1)\neq \pi ^{\ast }$ because $\theta ^{\prime }\neq \theta $. If $\hat{%
\sigma}(\theta ^{\prime },1)>\hat{\sigma}(\pi ^{\ast })$, then the optimal
assignment property requires $(\theta ^{\prime },1)$ to be assigned to $\pi
^{\ast }$, a contradiction. The only remaining possibility is that $\pi
(\theta ^{\prime },1)<\pi ^{\ast }$, which is only possible if $\theta
^{\prime }<\theta $. In this case, $\hat{\sigma}(\theta ^{\prime },1)\leq
1-\theta +\theta ^{\prime }$. By the optimal assignment property, $\sigma
(\theta ^{\prime },1)$ is weakly closer to $\hat{\sigma}(\theta ^{\prime
},1) $ than to $\hat{\sigma}(\pi ^{\ast })$. Therefore, $\sigma (\theta
^{\prime },1)\leq (1+1-\theta +\theta ^{\prime })/2$. By Remark \ref{remark
tit for tat}, $\sigma (\theta ^{\prime },0)\leq (\theta ^{\prime },1)$.
Therefore, $p_{\sigma }(\theta ^{\prime },1)\leq (2-\theta +\theta ^{\prime
})/2$, and therefore bounded away from $1$, where the gap is independent of $%
c$.

Finally, for every distinct $\theta ,\theta ^{\prime }\in \Theta ^{\ast }$, $%
\pi _{\theta }\neq \pi _{\theta ^{\prime }}$ because $\left\vert \hat{\sigma}%
(\pi _{\theta })-\hat{\sigma}(\pi _{\theta ^{\prime }})\right\vert
=\left\vert \theta -\theta ^{\prime }\right\vert $. Moreover, for generic $%
\Theta $, there cannot exist $\theta ^{\prime \prime }$ and $h$ such that $%
(\theta ^{\prime \prime },h)\in \pi _{\theta }$ and $(\theta ^{\prime \prime
},1-h)\in \pi _{\theta ^{\prime }}$, because that would require $\left\vert 
\hat{\sigma}(\pi _{\theta })-\hat{\sigma}(\pi _{\theta ^{\prime
}})\right\vert =\theta ^{\prime \prime }$, which is generically not the
case. It follows that for every $\theta \in \Theta ^{\ast }$, there is a
distinct additional payoff state $\theta ^{\prime }$, such that $p_{\sigma
}(\theta ^{\prime },1)$ is bounded away from one, independently of $c$. $%
\blacksquare $\bigskip

\noindent {\large Proposition \ref{prop monotone MLEQ}}

\noindent Suppose there is a monotone MLEQ $(\sigma ,\Pi )$ that sustains
positive cooperation rates in every payoff state. Then, as we saw in
previous results, $\hat{\sigma}(\theta ,1)-\hat{\sigma}(\theta ,0)=\theta $
for every $\theta $. Since $\max (\Theta )>\delta $, it follows that $\hat{%
\sigma}(\max (\Theta ),1)>\delta $. By the monotonicity requirement, $\hat{%
\sigma}(\theta ,1)>\delta $ for every $\theta $. Suppose $\pi (\theta
,1)\neq \pi (\theta ^{\prime },1)$ for some $\theta ,\theta ^{\prime }$.
Order the cells $\pi (\theta ,1)$ according to their $\hat{\sigma}$ value.
Monotonicity implies that if $\hat{\sigma}(\theta ^{\prime },1)>\hat{\sigma}%
(\theta ,1)$, then any member of $\pi (\theta ,1)$ is higher than any member
of $\pi (\theta ^{\prime },1)$. Consider two adjacent cells $\pi ,\pi
^{\prime }$. Merging them will not destroy monotonicity. Then, the
inequality (\ref{ineq ML optimal}) continues to be a necessary condition for
equilibrium:%
\[
c\leq \frac{p_{\sigma }(\pi (\theta ,1))p_{\sigma }(\pi (\theta ^{\prime
},1))}{p_{\sigma }(\pi (\theta ,1))+p_{\sigma }(\pi (\theta ^{\prime },1))}%
\left( \hat{\sigma}(\theta ,1)-\hat{\sigma}(\theta ^{\prime },1)\right) ^{2} 
\]%
However, the R.H.S. of this inequality is below $\frac{1}{2}(1-\delta )^{2}$%
, which is below $c$ if $\delta $ is sufficiently close to one.

It follows that when $\delta $ is large enough, all contingencies $(\theta
,1)$ belong to the same partition cell. Therefore, all partition cells $%
(\theta ,0)$ must belong to different cells, because $\hat{\sigma}(\theta
,1)-\hat{\sigma}(\theta ,0)=\theta $ for every $\theta $. Note that this
means $\hat{\sigma}(\theta ,0)$ is decreasing in $\theta $, which is
consistent with monotonicity. Thus, $\pi (\theta ,0)=\{(\theta ,0)\}$ ---
and hence $\hat{\sigma}(\theta ,0)=\sigma (\theta ,0)$ --- for every $\theta 
$. Consider two states $\theta $ and $\theta ^{\prime }$ that have adjacent $%
\sigma (\theta ,0)$ and $\sigma (\theta ^{\prime },0)$. Merging $\{(\theta
,0)\}$ and $\{(\theta ^{\prime },0)\}$ into one cell would not destroy
monotonicity. Therefore, (\ref{ineq ML optimal}) continues to be a necessary
condition for equilibrium:%
\begin{equation}
c\leq \frac{p_{\sigma }(\theta ,0)p_{\sigma }(\theta ^{\prime },0)}{%
p_{\sigma }(\theta ,0)+p_{\sigma }(\theta ^{\prime },0)}\left( \sigma
(\theta ,0)-\sigma (\theta ^{\prime },0)\right) ^{2}
\label{ineq condition monotone}
\end{equation}%
Since $\hat{\sigma}(\theta ,1)>\delta $ for every $\theta $, $\sigma (\theta
,1)>\delta $ for at least one state $\theta $. As $\delta \rightarrow 1$, $%
p_{\sigma }(\theta ,0)\rightarrow 0$. It follows that the R.H.S. is below $c$
if $\delta $ is sufficiently close to one, contradicting the assumption that 
$(\sigma ,\Pi )$ is a monotone MLEQ. $\blacksquare $\bigskip \bigskip

\noindent {\LARGE Appendix II: The Objective Meaning of }$c$\medskip

\noindent The model in this paper treats $c$ as a primitive, as if players
have an intrinsic taste for simple beliefs. However, under the ML
interpretation, we should view ML-optimality --- and the role that $c$ plays
in it --- as a reduced-form formalization of an underlying bias-variance
trade-off. This trade-off would arise in a more elaborate model in which
players do not observe $\sigma $ directly, but instead learn about it from a
noisy sample.

Specifically, suppose that $n=1$ and that for every $h=0,1$, players observe 
$x(h)=\sigma (h)+\varepsilon (h)$, where $\varepsilon (h)$ is an independent
noise term with mean zero and variance $v/p_{\sigma }(h)$; $v>0$ is a
constant. The basic model would correspond to the case in which $v=0$. The
assumption that the variance is inversely proportional to $p_{\sigma }(h)$
is in the spirit of Danenberg and Spiegler (2024): Players obtain a
representative finite sample drawn from the ergodic distribution over
contingencies, such that the number of observations about a contingency is
proportional to its frequency.

Fix a partition $\Pi $ of $H=\{0,1\}$. For every cell $\pi \in \Pi $, define 
$\hat{\sigma}(h)$ as the expected value of $x$ in $\pi (h)$. As before, the
MSPE it induces is $E(\hat{\sigma}(h)-\sigma (h))^{2}$. Then, the MSPE\
induced by the fine partition is%
\[
p_{\sigma }(0)\cdot \frac{v}{p_{\sigma }(0)}+p_{\sigma }(1)\cdot \frac{v}{%
p_{\sigma }(1)}=2v 
\]%
By comparison, the MSPE induced by the degenerate, coarse partition is%
\begin{eqnarray*}
&&p_{\sigma }(0)p_{\sigma }(1)[\sigma (1)-\sigma (0)]^{2}+(p_{\sigma
}(0))^{2}\cdot \frac{v}{p_{\sigma }(0)}+(p_{\sigma }(1))^{2}\cdot \frac{v}{%
p_{\sigma }(1)} \\
&=&p_{\sigma }(0)p_{\sigma }(1)[\sigma (1)-\sigma (0)]^{2}+v
\end{eqnarray*}

It follows that even if there is no intrinsic preference for simple beliefs
and ML-optimality is entirely based on minimizing MSPE, the fine partition
is ML-optimal if%
\[
p_{\sigma }(0)p_{\sigma }(1)[\sigma (1)-\sigma (0)]^{2}\geq v 
\]%
This is the same criterion as in the basic model, except that the role of $c$
in the basic model is now played by the noise variance constant $v$.

This little exercise cannot be turned into a full-fledged \textquotedblleft
foundation\textquotedblright\ for MLEQ. Apart from considerations of
tractability and generalizability, the fundamental difficulty is that if
players form beliefs according to a finite sample with smooth noise, they
will almost always have a strict best-replying action to their sample-based
belief. This action is independent of the observed history. Yet,
history-dependent behavior is crucial for sustaining cooperation in the OLG
trust game. This is a limitation of the game as a vehicle for exploring the
problem of long-run cooperation under ML-generated beliefs.

Nevertheless, the exercise gives a sense of how we may want to interpret the
inequality in Proposition \ref{prop bound}. It essentially a says that as
the number $m$ of payoff states that exhibit cooperation grows, the variance
of the noise with which players observe $\sigma $ should decrease at a rate
of approximately $1/m^{3}$.


\bigskip \bigskip

\begin{thebibliography}{99}
\bibitem{} Agrawal, A., J. Gans, and A. Goldfarb (2022). Prediction
machines, updated and expanded: The simple economics of artificial
intelligence. Harvard Business Press.

\bibitem{} Banchio, M. and G. Mantegazza (2022). Artificial intelligence and
spontaneous collusion. arXiv preprint arXiv:2202.05946.

\bibitem{} Basu, A., H. Shioya, and C. Park (2011). Statistical inference:
The minimum distance approach. CRC press.

\bibitem{} Bendor, J., D. Mookherjee, and D. Ray (2001). Reinforcement
learning in repeated interaction games. The BE Journal of Theoretical
Economics 1, 20011004.

\bibitem{} Brown, Z. and A. MacKay (2023). Competition in pricing
algorithms. American Economic Journal: Microeconomics 15, 109-156.

\bibitem{} Calvano, E., G. Calzolari, V. Denicolo, and S. Pastorello (2020).
Artificial intelligence, algorithmic pricing, and collusion. American
Economic Review, 110, 3267-3297.

\bibitem{} Danenberg, T. and R. Spiegler (2024). A Representative-Sampling
Model of Stochastic Choice. Journal of Political Economy: Microeconomics,
forthcoming.

\bibitem{} Eliaz, K. (2003). Nash equilibrium when players account for the
complexity of their forecasts. Games and Economic Behavior 44, 286-310.

\bibitem{} Eliaz, K. and R. Spiegler (2019). The model selection curse.
American Economic Review: Insights, 1, 127-140.

\bibitem{} Eliaz, K. and R. Spiegler (2022). On Incentive-Compatible
Estimators. Games and Economic Behavior 132, 204-220.

\bibitem{} Hansen, K., K. Misra, and M. Pai (2021). Algorithmic collusion:
Supra-competitive prices via independent algorithms. Marketing Science, 40,
1-12.

\bibitem{} Hastie, T., R. Tibshirani, and J. Friedman (2009). The elements
of statistical learning: Data mining, inference, and prediction. Springer.

\bibitem{} Jehiel, P. (2005). Analogy-based expectation equilibrium, Journal
of Economic theory 123, 81-104.

\bibitem{} Jehiel, P. and E. Mohlin (2024). Categorization in games: A
bias-variance perspective. Working paper.

\bibitem{} Jehiel, P. and G. Weber (2024). Endogenous clustering and
analogy-based expectation equilibrium. Working paper.

\bibitem{} Kaufman, L. and P. Rousseeuw (1990). Finding groups in data: An
introduction to cluster analysis. John Wiley \& Sons.

\bibitem{} Levine, S., A. Kumar, A., G. Tucker, and J. Fu (2020). Offline
reinforcement learning: Tutorial, review, and perspectives on open problems.
arXiv preprint arXiv:2005.01643.

\bibitem{} Mohlin, E. (2014). Optimal categorization. Journal of Economic
Theory 152, 356-381.

\bibitem{} Rubinstein, A. (1986). Finite automata play the repeated
prisoner's dilemma. Journal of economic theory 39, 83-96.

\bibitem{} Rubinstein, A. (1998). Modeling bounded rationality. MIT Press.

\bibitem{} Simon, H. (1982). Models of bounded rationality. MIT Press.

\bibitem{} Spiegler, R. (2002). Equilibrium in justifiable strategies: A
model of reason-based choice in extensive-form games. Review of Economic
Studies 69, 691-706.

\bibitem{} Spiegler, R. (2004). Simplicity of beliefs and delay tactics in a
concession game. Games and Economic Behavior 47, 200-220.

\bibitem{} Spiegler, R. (2005). Testing threats in repeated games. Journal
of Economic Theory 121, 214-235.

\bibitem{} Sutton, R and A. Barto (1998). Reinforcement learning: An
introduction. MIT press.

\bibitem{} Waizmann, S. (2024). AI in Action: Algorithmic Learning with
Strategic Consumers. Working paper.\bigskip \bigskip
\end{thebibliography}
\end{document}